\documentclass[11pt, a4paper]{article}
\usepackage{jheparxiv}
\usepackage{amsmath}
\usepackage{amsfonts}
\usepackage{amssymb}
\usepackage{latexsym}
\usepackage{mathrsfs}
\usepackage{graphicx}
\usepackage{color}
\usepackage{slashed}
\usepackage{twistor}
\usepackage[all]{xy}
\usepackage{hyperref}
\hypersetup{
  colorlinks = true,
  citecolor = blue,
  urlcolor = blue
}

\newcommand{{\bfr}}{\mbox{\boldmath$r$\unboldmath}}
\newcommand{{\bfv}}{\mbox{\boldmath$v$\unboldmath}}
\newcommand{{\bff}}{\mbox{\boldmath$f$\unboldmath}}

\newcommand{{\bfA}}{\mbox{\boldmath$A$\unboldmath}}
\newcommand{{\bfchi}}{\mbox{\boldmath$\chi$\unboldmath}}
\newcommand{\gradv}{\boldsymbol{\nabla}}
\newcommand{{\bfk}}{\mbox{\boldmath$k$\unboldmath}}

\newcommand{{\cY}}{\mbox{\boldmath${\cal Y}$\unboldmath}}
\def\v#1{{\bf#1}}

\title{The Helmholtz theorem and retarded fields}

\author{Ricardo Heras}

\affiliation{Department of Physics and Astronomy \\
        University College London, London, WC1E 6BT, UK\\
        }

\emailAdd{ricardo.heras.13@ucl.ac.uk}

\abstract{Textbooks frequently use the Helmholtz theorem to derive expressions for the electrostatic and magnetostatic fields but they do not usually apply this theorem to derive expressions for the time-dependent electric and magnetic fields, even when there is no formal objection to doing so because the proof of the theorem does not involve time derivatives but only spatial derivatives. Here we address the question as to whether the Helmholtz theorem is useful to derive expressions for the fields of Maxwell's equations. We show that when this theorem is applied to Maxwell's equations we obtain instantaneous expressions of the electric and magnetic fields, which are formally correct but of little practical usefulness. We then discuss two generalizations of the theorem which are shown to be useful to derive the retarded fields. }

\keywords{Helmholtz theorem, Maxwell's equations, retarded fields}

\begin{document}

\maketitle

\section{Introduction}
 \renewcommand{\theequation}{\arabic{equation}}
 \setcounter{equation}{0}

The mathematical formalism of electrostatics and magnetostatics is based on the Helmholtz theorem for time-independent vector fields \cite{1,2}.
According to this theorem any vector field $\v F(\v r)$ that goes to zero faster than $1/r$ as $r\to \infty$ can be expressed as\footnote[1]{The Helmholtz theorem can also be formulated as a uniqueness theorem which states that if the divergence $D(\v r)$  and the curl $\v C(\v r)$ of a vector field are specified, and if they both go to zero faster than $1/r^2$ as $r\to\infty$, then $\v F(\v r)$ is given uniquely by $\v F=-\gradv U +\gradv \times \v W$, where $U(\v r)=\int d^3r'  D(\v r')/(4\pi R)$ and $\v W(\v r)=\int d^3r'  \v C(\v r')/(4\pi R)$. See \cite{1} and \cite{2}.}
\begin{eqnarray}
\v F (\v r)= -\gradv\! \int \frac{\gradv' \cdot \v F ( \v r') }{4\pi R} \, d^3r'  + \gradv \times \int \!\frac{\gradv' \times \v F ( \v r') }{4\pi R}\, d^3r',
\end{eqnarray}

where the integrals are extended over all space, $R=|\v r-\v r'|$ with $\v r$ being the field point and $\v r'$ the source point. This theorem naturally applies to electrostatics and magnetostatics. If $\v F=\v E,$  $\gradv \cdot \v E=\rho/\epsilon_0$ and $\gradv \times \v E=0$ then (1) yields the electrostatic field $\v E(\v r)$ in SI units produced by the charge density $\rho(\v r)$:
\begin{eqnarray}
\v E(\v r)= -\frac{1}{4\pi\epsilon_0}\gradv \!\int \!\frac{\rho( \v r') }{ R} \, d^3r',
\end{eqnarray}
where $\epsilon_0$ is the permittivity constant of vacuum. If
$\v F=\v B,$  $\gradv \cdot \v B=0$ and $\gradv \times \v B=\mu_0\v J$ then (1) yields the magnetostatic field $\v B(\v r)$ in SI units produced by the current density $\v J(\v r)$:
\begin{eqnarray}
\v B (\v r)= \frac{\mu_0}{4\pi}\gradv \times\!\int\! \frac{\v J( \v r') }{R}\, d^3r',
\end{eqnarray}
where $\mu_0$ is the permeability constant of vacuum. The constants $\epsilon_0$ and $\mu_0$ satisfy  $\epsilon_0\mu_0=1/c^2$, where $c$ is the speed of light in vacuum. The Helmholtz theorem is then a powerful tool to find the electrostatic and magnetostatic fields of prescribed charge and current densities.

The question naturally arises: Is the Helmholtz theorem also useful to find expressions for the time-dependent electric and magnetic fields of Maxwell's equations? In particular, is it useful to find the retarded fields? Of course, this question is not a new one and has been addressed in one form or another by several authors \cite{3,4,5,6,7,8,9,10,11}, but given the pedagogical importance of the Helmholtz theorem in the time-independent regime of Maxwell's equations, it is worth reviewing again wether or not this theorem is equally useful in the time-dependent regime of these equations.

In section 2 we make the replacements $\v F(\v r)\!\to\! \v F(\v r,t)$ and $\v F(\v r')\!\to\! \v F(\v r',t)$ in equation (1) to obtain an instantaneous form of this equation. We apply this instantaneous form of the Helmholtz theorem to Maxwell's equations, obtaining instantaneous expressions for the electric and magnetic fields, which are formally correct but of little practical usefulness because they do not express the fields only in terms of their true sources: the charge and current densities. In section 3 we briefly mention some generalizations of the Helmholtz theorem that have been presented in the literature over the years.
We then focus our attention on two extensions of the theorem \cite{4,11} which are shown to be useful to derive the retarded fields. In section 4 we discuss the first extension of the theorem which we will call the \emph{causal} Helmholtz theorem because it expresses the vector field in terms of its sources evaluated at the retarded time \cite{4}. This causal Helmholtz theorem can  be presented in an undergraduate electrodynamics course.
We apply this causal form of the theorem to Maxwell's equations and obtain the retarded fields.
In section 5 we discuss the second extension of the theorem which we will call the causal Helmholtz theorem for antisymmetric tensor fields \cite{11,12}. It expresses an antisymmetric tensor field defined in the Minkowski space-time in terms of its sources multiplied by the retarded Green function of the wave equation. This generalization of the theorem is suitable for a graduate electrodynamics course. We apply this generalized theorem to the covariant form of Maxwell's equations and obtain the retarded electromagnetic field tensor, which involves the retarded electric and magnetic fields. In section 6 we apply both the instantaneous and causal forms of the Helmholtz theorem to Maxwell's equations expressed in terms of potentials, obtaining the corresponding instantaneous and retarded expressions of the vector potential in terms of their sources. In section 7 we present our conclusions. In Appendix A we prove the causal Helmholtz theorem and in Appendix B we demonstrate the causal Helmholtz theorem for antisymmetric tensor fields.

\section{Application of the theorem to time-dependent electric and magnetic fields}
 \renewcommand{\theequation}{\arabic{equation}}
 \setcounter{equation}{3}

The derivation of (1) does not involve time. This means that (1) can be applied to time-dependent vector fields as well.
Accordingly, we can make the replacements $\v F(\v r)\!\to\! \v F(\v r,t)$ and $\v F(\v r')\!\to\! \v F(\v r',t)$ in equation (1) to obtain the extended theorem \cite{4}:
\begin{eqnarray}
\v F (\v r, t)\!= -\gradv\!\int \frac{\gradv'\! \cdot \v F ( \v r', t) }{4\pi R} \, d^3r' +\! \gradv \! \times \! \int \frac{\gradv'\! \times \v F ( \v r', t) }{4\pi R}\, d^3r'.
\end{eqnarray}
Because (4) expresses the field $\v F (\v r, t)$ in terms of its sources $\nabla\! \cdot \v F ( \v r, t)$ and $\nabla\! \times \v F ( \v r, t)$ evaluated at the same instant of time $t$, we will call (4) the \emph{instantaneous} form of the Helmholtz theorem.

We can directly apply (4) to Maxwell's equations with sources in
vacuum. If $\v F=\v E$, $\gradv \cdot \v E=\rho/\epsilon_0$ and $\gradv \times\v E=-\partial \v B/\partial t$ then (4) gives
\begin{eqnarray}
\v E(\v r,t)= -\frac{1}{4\pi\epsilon_0}\gradv \!\int \frac{\rho(\v r',t)}{R}d^3r' -\frac{1}{4\pi}\gradv \times\!
\int\! \frac{\partial\v B(\v r',t)/\partial t}{R} d^3r'.
\end{eqnarray}
On the other hand, if $\v F=\v B$, $\gradv \cdot \v B=0$ and $\gradv \times\v B=\mu_0\v J +\epsilon_0\mu_0\partial \v E/\partial t$ then (4) yields
\begin{eqnarray}
\v B (\v r, t)=\frac{\mu_0}{4\pi} \gradv \times \! \int\!\frac{ \v J ( \v r', t) }{R}\, d^3r' + \frac{1}{4\pi c^2}
 \,\gradv \times\!
\int \!\frac{\partial \v E(\v r',t)/\partial t}{R} d^3r'.
\end{eqnarray}

Equations (5) and (6) are formally correct but they exhibit two practical and conceptual inconveniences: (i) To determine $\v E$ we need to specify $\rho$ and $\partial\v B/\partial t$  and to find $\v B$ we need to specify $\v J$ and $\partial\v E/\partial t$. However, the values of $\partial\v B/\partial t$ and $\partial\v E/\partial t$ are not generally known and therefore (5) and (6) are of limited usefulness; (ii) The field $\v E$ in (5)
is instantaneously related with their ``sources'' $\rho$ and $\partial\v B/\partial t$. Similarly, the field $\v B$ in (6)
is instantaneously related with their ``sources'' $\v J$ and $\partial\v E/\partial t$. Instantaneous connections between fields and their sources disagree with the experimentally verified causality property according to which the occurrence of the causes (the sources) precedes in time to the occurrence of the effects (the fields).

We introduce the operator $\gradv \times$ into the second integral of (5),
use $\gradv \times\v B=\mu_0\v J +\epsilon_0\mu_0\partial \v E/\partial t$ and perform an integration by parts to obtain the expression
\begin{eqnarray}
\v E(\v r,t)\!=-\frac{1}{4\pi \epsilon_0} \gradv \! \int\! \frac{\rho(\v r',t)}{R}d^3r'\! -\!\frac{\mu_0}{4\pi}\frac{\partial}{\partial t}\!
\int\!\frac{ \v J ( \v r', t)+ \epsilon_0 \partial \v E(\v r', t)/ \partial t}{ R}d^3r'.
\end{eqnarray}
By a similar procedure, equation (6)  and $\gradv \times\v E=-\partial \v B/\partial t$ yield  the equation
\begin{eqnarray}
\v B (\v r, t)\!= \frac{\mu_0}{4\pi}\gradv \!\times \! \int\!\frac{ \v J ( \v r', t) }{R}\, d^3r' - \frac{1}{ 4\pi c^2}\frac{\partial^2}{\partial t^2}\!\int \frac{\v B(\v r',t)}{R} d^3r'.
\end{eqnarray}
Both (7) and (8) are again instantaneous equations which do not express the fields $\v E$ and $\v B$ only in terms of their true sources: $\rho$ and $\v J$. However, if we apply the Laplacian operator $\gradv^2$ to (7) and (8), use $\gradv^2 (1/R)= -4\pi \delta(\v r-\v r')$, where $\delta$ is the Dirac delta function, and integrate over all space then we obtain the familiar wave equations
\begin{eqnarray}
\gradv^2\v E-\frac{1}{c^2}\frac{\partial^2 \v E}{\partial t^2}=\frac{1}{\epsilon_0} \gradv \rho + \mu_0 \frac{\partial \v J}{\partial t},\quad
\gradv^2\v B-\frac{1}{c^2}\frac{\partial^2 \v B}{\partial t^2}=-\mu_0\gradv \times \v J,
\end{eqnarray}
which can be solved to obtain the standard retarded fields:
\begin{eqnarray}
\v E=\frac{1}{4\pi\epsilon_0}\!\int \frac{[-\gradv'\rho-(1/c^2)\partial\v J/\partial t]}{R}  d^3r',\quad
\v B= \frac{\mu_0}{4\pi} \!\int \frac{[\gradv'\times\v J]}{R}d^3r',
\end{eqnarray}
where the square bracket $[\;\;]$ means that the enclosed quantity is to be evaluated at the retarded time $t'=t-R/c$. By applying the instantaneous form of the Helmholtz theorem to Maxwell's equations and following an indirect procedure we have arrived at the retarded fields. Nevertheless, this indirect procedure is pedagogically uninteresting because a direct manipulation of Maxwell's equations yields the corresponding wave equations in (9) which are then integrated to obtain the retarded fields displayed in (10) without the necessity of applying the Helmholtz theorem.

\section{Generalizations of the Helmholtz theorem}
Over the years some generalizations of the Helmholtz theorem have been formulated and applied to Maxwell's equations. In 1970 Hauser \cite{13} extended the Helmholtz theorem to four-vectors in the Minkowski space-time. This extension of the theorem is similar to that of three-vectors: Any four-vector is expressible as the sum of a solenoidal part and an irrotational part. In this case, however, the curl of a four-vector is an antisymmetric tensor field. A corollary of this theorem states that an antisymmetric tensor field vanishing sufficiently rapidly at spatial infinity is determined by specifying its divergence and the divergence of its dual. An explicit formula for this corollary involving the retarded Green function of the wave equation was introduced by Kobe \cite{11} in 1986. This corollary can naturally be considered as another generalization of the Helmholtz theorem for antisymmetric tensor fields. Kobe applied this generalization of the theorem to the covariant form of Maxwell's equations with magnetic monopoles. A short proof of this generalized theorem was given by Heras \cite{12} in 1990.

On the other hand, Kapu\'scik \cite{14} suggested in 1986 a generalization of the Helmholtz theorem for two time-dependent vector fields. According to this generalization, for each pair of time-dependent vector fields there exist two scalar potentials and two vector potentials with which the two vector fields can be constructed. Interestingly, when referring to the standard Helmholtz theorem, Kapu\'scik \cite{14} pointed out:
``There does not exist any simple generalization of this theorem for time-dependent vector fields $\v V(\v r, t).$'' He concluded that: ``For time-dependent
vector fields we have, therefore, two vector fields $\partial\v V/\partial t$  and rot $\v V$ and one scalar
field div $\v V$ to represent a single vector field $\v V(\v r,t)$ and there is no simple and useful
way to do that.'' Nevertheless, rewriting a result of Macquistan's book \cite{15}, Heras \cite{4} formulated in 1990 a useful extension of the Helmholtz theorem which states that a time-dependent vector field is determined by specifying  its curl, div and time derivative, being all of these quantities
evaluated at the retarded time. The expression that proves this generalization of the Helmholtz theorem leads to the retarded fields via a slightly indirect procedure as we will see in the next section. An equivalent form of this generalized theorem using the Green function of the wave equation was also formulated by Heras \cite{16} in 1995 to derive the time-dependent generalizations of the Coulomb and Biot-Savart laws in the form given by Jefimenko \cite{17}. In 1999 and 2009 Woodside \cite{18, 10} discussed some generalizations of the Helmholtz theorem for four-vectors. More recently, in 2016 Chubykalo et al \cite{19} have also reviewed the Helhmoltz theorem and its extension to four-vectors.

\section{The causal Helmholtz theorem}
 \renewcommand{\theequation}{\arabic{equation}}
 \setcounter{equation}{10}

In this section we will review the extension of the Helmholtz theorem, which leads to the retarded fields \cite{4,16,10,18,19}. Let us define a causal vector field as a time-dependent vector field $\v F(\v r,t)$ bounded in time whose sources are evaluated at the retarded time $t'=t-R/c$. The \emph{causal} Helmholtz theorem can be formulated as follows: any causal vector field $\v F(\v r,t)$  that goes to zero faster than $1/r$ as $r\to \infty$ can be expressed as \cite{4}:
\begin{eqnarray}
\v F = -\gradv \!\int \frac{[\gradv' \cdot \v F]  }{4\pi R} \, d^3r' + \gradv \times\!  \int \frac{[\gradv' \times \v F]}{4\pi R}\, d^3r' +\frac{1}{c^2}\frac{\partial}{\partial t}\!\int \frac{[\partial \v F/\partial t]}{4\pi R} \, d^3r',
\end{eqnarray}
where the square bracket $[\;\;]$ means that the enclosed quantity is to be evaluated at the retarded time $t'=t-R/c$. As may be seen, the sources of the field $\v F$ are its divergence, curl and time derivative, all of them evaluated at the retarded time. In the Appendix A we derive (11). As expected, the causal Helmholtz theorem naturally applies to Maxwell's equations.

If we write $\v F\!=\!\v E$ in (11) and use $\gradv \cdot\v E\!=\!\rho/\epsilon_0$, $\gradv \times \v E\!=\!-\partial \v B/\partial t$, and $\partial \v E/\partial t\!=\!c^2\gradv \times \v B\!-\!\mu_0c^2\v J$ then we obtain the equation
\begin{equation}
 \v E =-\frac{1}{4\pi\epsilon_0}\gradv \!\int\! \frac{[\rho]}{ R} d^3r'-\frac{\mu_0}{4\pi}\frac{\partial}{\partial t}\!\int\! \frac{[\v J]}{R}d^3r'  - \gradv \times\!\int\!\frac{[\partial\v B/\partial t]}{4\pi R}d^3r' +\frac{\partial}{\partial t}
\!\int\! \frac{[ \gradv'\times \v B]}{4\pi R}d^3r'.
\end{equation}
Using the property $\partial [\v F]/\partial t = [\partial\v F/\partial t]$  and performing an integration by parts in which the corresponding surface term is seen to vanish at infinity, we obtain the result \cite{4}:
\begin{equation}
\gradv \times\!\int\!\frac{[\partial\v F/\partial t]}{R}d^3r'= \frac{\partial}{\partial t}
\!\int\! \frac{[ \gradv'\times \v F]}{R}d^3r'.
\end{equation}
Equations (12) and (13) yield the retarded electric field:
\begin{equation}
\v E =-\frac{1}{4\pi\epsilon_0}\gradv\!\int\! \frac{[\rho]}{ R} d^3r'-\frac{\mu_0}{4\pi}\frac{\partial}{\partial t}\!\int\! \frac{[\v J]}{R}d^3r'
\end{equation}

Similarly, if we let $\v F\!=\!\v B$ in (11) and use $\gradv \cdot \v B\!=\!0, \gradv\times\v B\!=\!\mu_0\v J +\epsilon_0\mu_0\partial \v E/\partial t$ and $\partial \v B/\partial t\!=\!-\gradv \times\v E$ then we obtain
\begin{equation}
\v B =\gradv \times \int\! \frac{[\mu_0\v J]}{4\pi R}d^3r'+\frac{1}{c^2}\gradv \times\!\int\!\frac{[\partial\v E/\partial t]}{R}d^3r'
-\frac{1}{c^2}\frac{\partial}{\partial t}
\!\int\! \frac{[ \gradv'\times \v E]}{4\pi R}d^3r'.
\end{equation}
Using (13) in (15) we obtain the retarded magnetic field
\begin{equation}
\v B =\frac{\mu_0}{4\pi}\gradv \times\!\int\! \frac{[\v J]}{R}d^3r'.
\end{equation}
The fields $\v E$ and $\v B$ in (10) are equivalent to those in (14) and (16). This can be shown after performing an integration by parts. The causal Helmholtz theorem is then a useful tool to find the retarded fields. The derivation of (14) and (16) is suitable for an undergraduate electrodynamics course.

\section{The causal Helmholtz theorem for antisymmetric tensor fields}

 \renewcommand{\theequation}{\arabic{equation}}
 \setcounter{equation}{16}

In this section we will review the extension of the Helmholtz theorem suggested by Hauser \cite{13} and discussed by Heras \cite{12} and Kobe  \cite{11}. We will see that this extension of the theorem also leads to the retarded fields. Since this extension is formulated for antisymmetric tensor fields in the Minkowski space-time we need to introduce tensor notation. Greek indices $\mu, \nu, \kappa \ldots$ run from 0 to 3 and  Latin indices $i,j,k,\ldots$ run from 1 to 3. The summation convention on repeated indices is adopted. The signature of the metric is $(+,-,-,-).$ A point in space-time is denoted by $x^{\mu}\!=\!(x^0,x^i)\!=\!(ct,\v r).$ The four-gradient is defined as $\partial_\mu\!=\![(1/c)\partial/\partial t, \nabla].$ The Kronecker delta in space-time is denoted by $\delta^\mu_\nu$. The totally antisymmetric four-dimensional Levi-Civita symbol reads $\varepsilon^{\mu\nu\alpha\beta}$ with $\varepsilon^{0123}\!=\!1$ and $\varepsilon^{ijk}$ is the totally antisymmetric three-dimensional Levi-Civita symbol with $\varepsilon^{123}\! = \!1$. Using the four-gradient and the Kronecker delta we can construct the useful operator \cite{12}: $\partial^{\mu\nu}_{\;\;\,\lambda}$, which is antisymmetric in $\mu$ and $\nu$. This operator and its associated dual operator are defined as
\begin{eqnarray}
\partial^{\mu\nu}_{\;\;\;\lambda}=\delta^\nu_\lambda\partial^\mu-\delta^\mu_\lambda\partial^\nu,   \quad   ^{\ast}\!\partial^{\mu\nu}_{\;\;\;\lambda}=\frac{1}{2}\varepsilon^{\mu\nu\alpha\beta}\partial_{\alpha\beta \lambda}=\varepsilon^{\mu\nu\alpha}_{\quad\lambda}\partial_{\alpha}.
\end{eqnarray}
These operators are seen to satisfy the properties
\begin{eqnarray}
\partial_\mu\partial^{\mu\nu}_{\;\;\;\lambda}=\delta^\nu_\lambda\partial_\mu\partial^\mu-\partial_\lambda\partial^\nu,   \quad   ^{\ast\,\ast}\!\partial^{\mu\nu}_{\;\;\;\lambda}= -\, \partial^{\mu\nu}_{\;\;\;\lambda},\quad
\partial_\mu\!\,^{\ast}\partial^{\mu\nu}_{\;\;\;\lambda}=0.
\end{eqnarray}
It follows that $\partial^{\mu\nu}_{\;\;\;\lambda}F^\lambda=\partial^\mu F^\nu-\partial^\nu F^\mu.$ Heras \cite{12} has demonstrated the following useful identity for an antisymmetric tensor field ${\cal F}^{\mu\nu}$,
\begin{eqnarray}
\partial_\alpha\partial^\alpha {\cal F}^{\mu\nu}\equiv\partial^{\mu\nu}_{\;\;\;\lambda}\big[\partial_{\alpha}{\cal F}^{\alpha\lambda}\big]\; -\; ^{\ast}\!\partial^{\mu\nu}_{\;\;\;\lambda}\big[\partial_{\alpha}{^{\ast}{\cal F}^{\alpha\lambda}}\big],
\end{eqnarray}
where $^{\ast}{\cal F}^{\mu\nu}$ is the dual of ${\cal F}^{\mu\nu}$, i.e., $^{\ast}{\cal F}^{\mu\nu}=(1/2)\varepsilon^{\mu\nu\kappa\lambda}{\cal F}_{\kappa\lambda}.$ For brevity, we are going to drop superscripts in the functional
dependencies of tensors. For example, we will write ${\cal B}^{\mu\nu}(x)$ to denote ${\cal B}^{\mu\nu}(x^\alpha)$. The identity (19) is the analogous antisymmetric-tensor of the double-curl identity $\gradv^2 \v F \!\equiv\! \gradv [\gradv \cdot \v F] - \gradv \times [\gradv \times \v F].$ We define a causal antisymmetric tensor field ${\cal F}^{\mu\nu}$ in the Minkowski space-time as a tensor whose sources are multiplied by the retarded Green function
$G=G(x,x')$ of the four-dimensional wave equation $\partial_\mu\partial^\mu G=\delta^{(4)}(x-x'),$
where $\delta^{(4)}(x\!-\!x')$ is the four-dimensional delta function. The function $G$ satisfies $\partial^\mu G=-\partial'^\mu G.$
An explicit form of the function $G$ reads $G=\delta\{ct'-ct+R \}/(4\pi R)$ \cite{10}.\footnote[2]{Notice that this form of $G$ is not an explicit Lorentz-invariant function. An equivalent form of $G$ which is Lorentz invariant reads $D_r(x,x')\!=\!\Theta(x_0\!-\!x'_0)\delta[(x\!-\!x')^2]/(2\pi),$ where $\Theta$ is the theta function. See \cite{20}.}

 The causal Helmholtz theorem for antisymmetric tensor fields in the Minkowski space-time states that a causal antisymmetric tensor field ${\cal F}^{\mu\nu}(x)$ vanishing sufficiently rapidly at spatial infinity can be expressed as \cite{12}:
\begin{eqnarray}
{\cal F}^{\mu\nu}(x)=\partial^{\mu\nu}_{\;\;\;\lambda}\!\int\!G\partial'_{\alpha}{\cal F}^{\alpha\lambda}(x')\,d^4x' -\; ^{\ast}\!\partial^{\mu\nu}_{\;\;\;\lambda}\!\int \!G\partial'_{\alpha} {^{\ast}{\cal F}^{\alpha\lambda}}(x')\,d^4x',
\end{eqnarray}
where $d^4x'$ is a volume element in the space-time and
the integrals are taken over all space-time. As may be seen, the sources of
the tensor field ${\cal F}^{\mu\nu}$ are its divergence $\partial_{\alpha}{\cal F}^{\alpha\lambda}$ and the divergence of its dual $\partial_{\alpha} {^{\ast}{\cal F}^{\alpha\lambda}}$. A proof of (20) will be presented in Appendix B. Equation (20) can directly be applied to the covariant form of Maxwell's equations
\begin{eqnarray}
\partial_\mu F^{\mu\nu}=\mu_0J^\nu,\quad \partial_\mu\,\!^*\!{F}^{\mu\nu}=0,
\end{eqnarray}
where $F^{\mu\nu}$ is the electromagnetic field tensor, $^{\ast}F^{\mu\nu}=(1/2)\varepsilon^{\mu\nu\kappa\lambda}F_{\kappa\lambda}$ is its associated dual and $J^\nu=(c\rho,\v J) $ is the four-current density.
If we write ${\cal F}^{\mu\nu}=F^{\mu\nu}$ and $^{\ast}{\cal F}^{\mu\nu}=\,^{\ast}{F}^{\mu\nu}$ in (20) and use (21) then
we obtain the retarded electromagnetic field
\begin{eqnarray}
F^{\mu\nu}(x)=\mu_0\,\partial^{\mu\nu}_{\;\;\;\lambda}\!\!\int\!GJ^\lambda(x')d^4x'.
\end{eqnarray}
In order to verify that (22) represents the vector form of the retarded fields, consider first the polar component $F^{i0}$:
\begin{align}
F^{i0}(x)&=\mu_0\,\partial^{i0}_{\;\;\;\lambda}\!\!\int\!GJ^\lambda(x')d^4x'
=\mu_0\,(\delta^0_\lambda\partial^i-\delta^i_\lambda\partial^0)\!\!\int\!GJ^\lambda(x')d^4x'\nonumber\\
&=\mu_0\,\partial^i\!\!\int\!GJ^0(x')d^4x'-\mu_0\,\partial^0\!\!\int\!GJ^i(x')d^4x'.
\end{align}
Inserting $F^{i0}\!=\!(\v E)^i/c,$  $J^0\!=\!c\rho,$  $J^i\!=\!(\v J)^i,$  $\partial^0\!=\!(1/c)\partial/\partial t,$  $\partial^i\!= \!-(\gradv)^i$ and  $G\!=\!\delta\{ct'\!-ct+R\}/(4\pi R)$ in (23) we obtain
\begin{eqnarray}
(\v E)^i(\v r, t)=-\frac{1}{\epsilon_0}(\gradv)^i\!\!\iint\!\frac{\delta\{t'\!-\!t\!+\!R/c\}}{4\pi R}\rho(\v r',t')d^3r'dt' \\\nonumber - \mu_0\frac{\partial}{\partial t} \!\!\iint\!\frac{\delta\{t'\!-\!t\!+\!R/c\}}{4\pi R}(\v J)^i(\v r',t')d^3r'dt',
\end{eqnarray}
where we have used the property $\delta\{cu\}=\delta\{u\}/c.$ Integration over time yields the retarded electric field expressed in index notation
\begin{equation}
(\v E)^i =-\frac{1}{4\pi\epsilon_0}\bigg(\gradv \!\int\! \frac{[\rho]}{R} d^3r'\bigg)^i-\frac{\mu_0}{4\pi}\bigg(\frac{\partial}{\partial t}\!\int\! \frac{[\v J]}{R}d^3r'\bigg)^i,
\end{equation}
where again the bracket $[\;\;]$ means that the enclosed quantity is to be evaluated at the retarded time $t'\!=\!t-R/c$. Following a similar procedure, we can show that the axial component $F^{ij}\!=\!-\varepsilon^{ijk}(\v B)_k$ yields the retarded magnetic field expressed in index notation
\begin{equation}
(\v B)^i =\frac{\mu_0}{4\pi}\bigg(\gradv \times\!\!\int\! \frac{[\v J]}{R}d^3r'\bigg)^i.
\end{equation}
The causal Helmholtz theorem for antisymmetric tensor fields is then a useful tool to find the retarded fields of prescribed charge and current densities. This generalization of the Helmholtz theorem involves the retarded Green function of the wave equation in the four-dimensional space and therefore it is suitable for a graduate electrodynamics course.

\section{Instantaneous and causal forms of the Helmholtz theorems applied to potentials}
 \renewcommand{\theequation}{\arabic{equation}}
 \setcounter{equation}{26}

In this section we will apply both the instantaneous and causal forms of the Helmholtz theorem given in (4) and (11) to Maxwell's equations expressed in terms of potentials and as a result we will obtain the corresponding instantaneous and retarded expressions of the vector potential $\v A$ in terms of their sources.

We first note that the homogeneous Maxwell's equations $\gradv \cdot\v B=0$ and $\gradv \times\v E + \partial\v B/\partial t=0$ imply the existence of the potentials $\Phi(\v r,t)$ and $\v A(\v r,t)$ such that
 \begin{eqnarray}
-\gradv \Phi(\v r,t)-\frac{\partial \v A(\v r,t)}{\partial t}=\v E(\v r,t), \quad \gradv \times\v A(\v r,t)=\v B(\v r,t).
\end{eqnarray}
We apply the instantaneous form of the Helmholtz theorem to the vector potential,
\begin{eqnarray}
\v A (\v r, t)\!= -\gradv \!\int \frac{\gradv'\! \cdot \v A ( \v r', t) }{4\pi R} \, d^3r' + \gradv \times \! \int \frac{\gradv'\! \times \v A ( \v r', t) }{4\pi R}\, d^3r'.
\end{eqnarray}

To determine the potential $\v A$ we need to specify $\gradv \cdot \v A$ and  $\gradv \times \v A$.  The latter quantity is already given by the second equation in (27). But we are free to specify $\gradv \cdot \v A$. A criterion to fix the value of $\gradv \cdot \v A$ must take into account two aspects: (i) The first one deals with the question of what are the sources of the magnetic field $\v B$.  If we consider that both the current density $\v J$ and the displacement current $\epsilon_0\partial \v E/\partial t$ are
 the sources of $\v B$ then the potential $\v A$ should be generally expressed in terms of these currents on account of $\v B=\gradv\times \v A$; (ii) The specification $\gradv \cdot \v A$ should be such that the potential $\v A$ is uniquely determined by their specified currents and appropriate boundary conditions. Both aspects are fulfilled if we assume that $\v A$  is solenoidal
 \begin{eqnarray}
\gradv \cdot\v A(\v r, t)=0.
\end{eqnarray}
Using (29) and the second equation of (27) in (28) we get
\begin{eqnarray}
\v A (\v r, t)= \gradv\! \times \! \int \frac{\v B( \v r', t) }{4\pi R}\, d^3r'.
\end{eqnarray}
Evidently, this expression for $\v A$ satisfies (29) back.
Introducing the operator $\gradv \times$ into the integral of (30), performing an integration by parts and using the Ampere-Maxwell law $\gradv\times\v B=\mu_0\v J +\epsilon_0\mu_0\partial \v E/\partial t$ we obtain the instantaneous vector potential
\begin{eqnarray}
\v A (\v r, t)= \frac{\mu_0}{4\pi}\int\!\frac{ \v J ( \v r', t)+ \epsilon_0 \partial \v E(\v r', t)/ \partial t}{ R}d^3r'.
\end{eqnarray}

The first equation in (27) and Gauss's law $\gradv\cdot\v E\!=\!\rho/\epsilon_0$ yield $\gradv^2 \Phi + \partial (\gradv \cdot \v A)/\partial t  =-\rho/\epsilon_0$, which reduces to
$\gradv^2 \Phi =-\rho/\epsilon_0$ on account of (29). The solution of this Poisson equation is
 \begin{eqnarray}
\Phi(\v r, t)=\frac{1}{4\pi\epsilon_0}\int\! \frac{\rho(\v r',t)}{R}d^3r'.
\end{eqnarray}
When the potentials (31) and (32) are introduced in (27) we obtain the instantaneous expressions for the electric and magnetic fields given in (7) and (6).

We now apply the causal form of the Helmholtz theorem to the vector potential
\begin{eqnarray}
\v A = -\gradv \!\int \frac{[\gradv' \cdot \v A]  }{4\pi R} \, d^3r' + \gradv \times\!  \int \frac{[\gradv' \times \v A]}{4\pi R}\, d^3r' +\frac{1}{c^2}\frac{\partial}{\partial t}\!\int \frac{[\partial \v A/\partial t]}{4\pi R} \, d^3r',
\end{eqnarray}
In order to determine the potential $\v A$, we need to specify $\gradv \cdot \v A, \gradv \times \v A$ and  $\partial \v A/\partial t$. The curl of $\v A$ is already specified in the second equation in (27). Again we are free to specify $\gradv \cdot \v A$. To fix the value of $\gradv \cdot \v A$ we must consider two aspects: (i) that the true current of the magnetic field $\v B$ is the current density $\v J$. Since $\v B=\gradv \times \v A$ then the potential $\v A$ should be generally given in terms of the current $\v J$; (ii) the specification $\gradv \cdot \v A$ should be such that the potential $\v A$ is uniquely determined by their sources and appropriate boundary conditions. Both aspects are fulfilled
by the condition
\begin{eqnarray}
\gradv \cdot\v A(\v r, t)=-\frac{1}{c^2}\frac{\partial\Phi(\v r, t)}{\partial t}.
\end{eqnarray}
If we impose this condition and perform an integration by parts in the first term of (33) then we can link the resulting term with the third term of (33) and obtain the following expression
\begin{eqnarray}
\v A = \frac{1}{4\pi c^2}\frac{\partial}{\partial t}\!\int \frac{[\gradv'\Phi+\partial \v A/\partial t]  }{R} \, d^3r' + \gradv \times\!  \int \frac{[\v B]}{4\pi R}\, d^3r',
\end{eqnarray}
Introducing $\gradv \times$ into the second integral of
(35), performing an integration by parts and using the Ampere-Maxwell law $\gradv \times\v B=\mu_0\v J +\epsilon_0\mu_0\partial \v E/\partial t$ we get the equation
\begin{eqnarray}
\v A = \frac{1}{4\pi c^2}\frac{\partial}{\partial t}\!\int \frac{[\gradv'\Phi+\partial \v A/\partial t+\v E]  }{R} +\frac{\mu_0}{4\pi}\!\int\! \frac{[\v J]}{R}d^3r'.
\end{eqnarray}
When the first equation of (27) is used in (36) we arrive at the retarded vector potential
\begin{equation}
\v A =\frac{\mu_0}{4\pi}\!\int\! \frac{[\v J]}{R}d^3r'.
\end{equation}
On the other hand, the first equation in (27) and Gauss's law $\gradv \cdot\v E\!=\!\rho/\epsilon_0$ imply the equation $\gradv^2 \Phi + \partial (\gradv \cdot \v A)/\partial t  =-\rho/\epsilon_0$, which becomes
$\gradv^2 \Phi-(1/c^2)\partial^2 \Phi/\partial t^2=-\rho/\epsilon_0$ on account of (34). The retarded solution of this wave equation reads
 \begin{eqnarray}
\Phi=\frac{1}{4\pi\epsilon_0}\int\! \frac{[\rho]}{R}d^3r'.
\end{eqnarray}
When the potentials (37) and (38) are introduced in (27) we obtain the retarded fields in (14) and (16). Therefore, we have demonstrated two results: (I) The instantaneous Helmholtz theorem applied to the vector potential $\v A$ of Maxwell's equations implies an instantaneous expression for this potential in terms of the currents $\v J$ and $\epsilon_0\partial \v E/\partial t$ whenever the additional condition $\gradv\cdot\v A\!=\!0$ is assumed; (II) The causal Helmholtz theorem applied to the vector potential $\v A$ of Maxwell's equations implies a retarded expression for this potential in terms of the current $\v J$  whenever the additional condition $\gradv \cdot\v A+(1/c^2)\partial\Phi/\partial t\!=\!0$ is assumed. In the context of gauge invariance these two conditions are respectively called the Coulomb and Lorenz gauge conditions. However, we must emphasize that these conditions were imposed  here without explicitly considering the gauge invariance of Maxwell's equations.

\section{Conclusions}
We can draw some conclusions on our review on the Helmholtz theorem and its extensions as well as the applicability of these forms of the theorem to derive the retarded fields:
\begin{itemize}
  \item When the instantaneous form of the Helmholtz theorem given in (4) is applied to Maxwell's equations, we obtain instantaneous expressions of the electric and magnetic fields, which are formally correct but of little practical usefulness. This application of the theorem is not usually presented in textbooks but it is conceptually important to show that there is a dual description (one instantaneous and the other one retarded) of the electric and magnetic fields that satisfy Maxwell's equations for sources in vacuum.

   \item When the causal form of the Helmholtz theorem expressed in (11) is applied to Maxwell's equations and the property (13) used, we directly obtain expressions for the retarded electric and magnetic fields. This derivation is suitable for an undergraduate presentation of Maxwell's equations.

    \item When the causal form of the Helmholtz theorem for antisymmetric tensor fields expressed in (20) is applied to the covariant form of Maxwell's equations, we directly obtain an expression for the retarded electromagnetic field tensor, which involves the retarded electric and magnetic fields. This derivation is suitable for a graduate presentation of the covariant form of Maxwell's equations.

     \item  When the instantaneous form of the Helmholtz theorem is applied to the vector potential of Maxwell's equations and the Coulomb gauge condition is assumed, we obtain an instantaneous expression for this potential in terms of the conduction and displacement currents. When the causal Helmholtz theorem is applied to the vector potential of Maxwell's equations and the Lorenz gauge condition is assumed, we obtain a retarded expression for this potential in terms of the conduction current. These results are obtained without explicitly considering the gauge invariance of Maxwell's equations.
\end{itemize}


\appendix
 \renewcommand{\theequation}{A.\arabic{equation}}
 \setcounter{equation}{0}

\section{Derivation of equation (11)}
The starting point is the following identity for the time-dependent vector field $\v F(\v r,t)$:
\begin{equation}
\Box^2 \v F \equiv \gradv \big(\gradv \cdot \v F\big) - \gradv \times \big(\gradv \times \v F\big)-\frac{1}{c^2}\frac{\partial}{\partial t} \bigg( \frac{\partial \v F}{\partial t}\bigg),
\end{equation}
where $\Box^{2}\equiv\gradv^2-(1/c^2)\partial^2/\partial t^2$ is the d'Alembert operator. We evaluate (A.1) at the source point at the retarded time (using the retardation parentheses), multiply the resulting equation by $1/R$ and integrate over all space
\begin{equation}
\int\frac{[\Box'^2 \v F]}{R}d^3 r' \!=\!\! \int\frac{[\gradv'(\gradv' \cdot \v F)]}{R}d^3 r'
\!-\int\frac{[\gradv' \times (\gradv' \times \v F)]}{R}d^3 r'\!-\!\frac{1}{c^2}\! \int \frac{[\partial^2 \v F/\partial t^2]}{R}d^3 r'.
\end{equation}
By assuming that the field $\v F$ vanishes sufficiently rapidly at spatial infinity and performing successive integrations by parts in which the corresponding surface terms can be shown to vanish we obtain the results
\begin{eqnarray}
\int\frac{[\Box'^2 \v F]}{R}d^3 r' \!=\!\int\!\Box^2\bigg(\frac{[\v F]}{R}\bigg)d^3 r' = -4\pi\int[\v F]\delta(\v r-\v r')d^3 r'=&-4\pi\v F(\v r,t), \\
\int\frac{[\gradv'(\gradv' \cdot \v F)]}{R}d^3 r'\!=\!\gradv\!\!\int\frac{[\gradv' \cdot \v F]}{R}d^3 r',\\
\int\frac{[\gradv' \times (\gradv' \times \v F)]}{R}d^3 r'\!=\!\gradv \times \!\!\int\frac{[\gradv' \times \v F ]}{R}d^3 r',\\
\int \frac{[\partial^2 \v F/\partial t^2]}{R}d^3 r'\!=\!\frac{\partial}{\partial t}\!\!\int \frac{[\partial\v F/\partial t]}{R}d^3 r'.
\end{eqnarray}
In deriving (A.3) we have used the identity $\Box^2([\v F]/R)=-4\pi[\v F]\delta(\v r-\v r')$ which is proved in Ref. \cite{21}. The identity (A.5) is proved in Ref. \cite{22}. Using (A.3)-(A.6) into (A.2) we obtain the equation (11). We note that equation (11) can also be justified by taking its d'Alembertian, performing the corresponding integrations over all space and obtaining the identity (A.1).

\section{Derivation of equation (20)}
 \renewcommand{\theequation}{B.\arabic{equation}}
 \setcounter{equation}{0}

Consider the identity (19):
\begin{eqnarray}
\partial_\alpha\partial^\alpha {\cal F}^{\mu\nu}\equiv\partial^{\mu\nu}_{\;\;\;\lambda}\big[\partial_{\alpha}{\cal F}^{\alpha\lambda}\big]\; -\; ^{\ast}\!\partial^{\mu\nu}_{\;\;\;\lambda}\big[\partial_{\alpha}{^{\ast}{\cal F}^{\alpha\lambda}}\big],
\end{eqnarray}
We evaluate this identity at $x'$, multiply by the retarded Green function of the wave equation $G$ and integrate over all space-time,
\begin{eqnarray}
\int\! G\partial'_\alpha\partial'^\alpha {\cal F}^{\mu\nu}d^4x'=\int\! G \partial'^{\mu\nu}_{\;\;\;\lambda}\big[\partial'_{\alpha}{\cal F}^{\alpha\lambda}\big] d^4x'-\int\! G ^{\ast}\!\partial'^{\mu\nu}_{\;\;\;\lambda}\big[\partial'_{\alpha}{^{\ast}{\cal F}^{\alpha\lambda}}\big]   d^4x'.
\end{eqnarray}
Integrations by parts in (B.2) in which the corresponding surface terms vanish at space infinity (on account of ${\cal F}^{\mu\nu}$ vanishing sufficiently rapidly at spatial infinity) lead to the results
\begin{eqnarray}
\int\! G\partial'_\alpha\partial'^\alpha {\cal F}^{\mu\nu}d^4x'=\int\!{\cal F}^{\mu\nu} \partial'_\alpha\partial'^\alpha G d^4x'=\!\int\!{\cal F}^{\mu\nu}\delta^{(4)}(x-x') d^4x'=&{\cal F}^{\mu\nu}(x),\\
\int\! G\partial'^{\mu\nu}_{\;\;\;\lambda}\big[\partial'_{\alpha}{\cal F}^{\alpha\lambda}\big] d^4x'=\partial^{\mu\nu}_{\;\;\;\lambda}\!\int\! G \partial'_{\alpha}{\cal F}^{\alpha\lambda} d^4x',\\
\int\! G ^{\ast}\!\partial'^{\mu\nu}_{\;\;\;\lambda}\big[\partial'_{\alpha}{^{\ast}{\cal F}^{\alpha\lambda}}\big] d^4x'=\, ^{\ast}\!\partial^{\mu\nu}_{\;\;\;\lambda}\!\int\! G    \partial'_{\alpha}{^{\ast}{\cal F}^{\alpha\lambda}} d^4x'.
\end{eqnarray}
The use of (B.3)-(B.5) in (B.2) yields equation (20). Alternatively, (20) can be justified by simply taking its d'Alembertian, performing the corresponding integrations over all space-time and obtaining the identity displayed in (B.1).

\end{document}